\documentclass[superscriptaddress,prx,twocolumn, aps, 10pt, floatfix]{revtex4-2}
\usepackage{graphicx}
\usepackage{amsmath,amssymb,physics,dsfont}
\usepackage{color}
\usepackage{hyperref}
\usepackage{microtype}
\usepackage{braket}
\usepackage{booktabs}
\usepackage{defs-private}

\usepackage[dvipsnames]{xcolor}
\hypersetup{
    colorlinks=true,
    linkcolor=MidnightBlue,
    urlcolor=BrickRed,
    citecolor=Mulberry,
    pdftitle={Quantum subspace expansion in the presence of noise},
    pdfauthor={Jo\~{a}o C. Getelina, Prachi Sharma, Thomas Iadecola, Peter P. Orth, Yong-Xin Yao}
}

\begin{document}


\title{Quantum subspace expansion in the presence of hardware noise} 

\author{Jo\~{a}o C. Getelina}
\affiliation{Ames National Laboratory, U.S. Department of Energy, Ames, Iowa 50011, USA}

\author{Prachi Sharma}
\affiliation{Department of Physics, Saarland University, 66123 Saarbr\"ucken, Germany}

\author{Thomas Iadecola}
\email{iadecola@iastate.edu}
\affiliation{Ames National Laboratory, U.S. Department of Energy, Ames, Iowa 50011, USA}
\affiliation{Department of Physics and Astronomy, Iowa State University, Ames, Iowa 50011, USA}

\author{Peter P. Orth}
\email{peter.orth@uni-saarland.de}
\affiliation{Ames National Laboratory, U.S. Department of Energy, Ames, Iowa 50011, USA}
\affiliation{Department of Physics, Saarland University, 66123 Saarbr\"ucken, Germany}
\affiliation{Department of Physics and Astronomy, Iowa State University, Ames, Iowa 50011, USA}

\author{Yong-Xin Yao}
\email{ykent@iastate.edu}
\affiliation{Ames National Laboratory, U.S. Department of Energy, Ames, Iowa 50011, USA}
\affiliation{Department of Physics and Astronomy, Iowa State University, Ames, Iowa 50011, USA}

\begin{abstract}
    Finding ground state energies on current quantum processing units (QPUs) using algorithms like the variational quantum eigensolver (VQE) continues to pose challenges. Hardware noise severely affects both the expressivity and trainability of parametrized quantum circuits, limiting them to shallow depths in practice. Here, we demonstrate that both issues can be addressed by synergistically integrating VQE with a quantum subspace expansion, allowing for an optimal balance between quantum and classical computing capabilities and costs. We perform a systematic benchmark analysis of the iterative quantum-assisted eigensolver of [K. Bharti and T. Haug, Phys. Rev. A {\bf 104}, L050401 (2021)] in the presence of hardware noise. We determine ground state energies of 1D and 2D mixed-field Ising spin models on noisy simulators and on the IBM QPUs \texttt{ibmq\_quito} (5 qubits) and \texttt{ibmq\_guadalupe} (16 qubits). To maximize accuracy, we propose a suitable criterion to select the subspace basis vectors according to the trace of the noisy overlap matrix. Finally, we show how to systematically approach the exact solution by performing controlled quantum error mitigation based on probabilistic error reduction on the noisy backend \texttt{fake\_guadalupe}.
\end{abstract}

\maketitle

\section{Introduction}
The rapid advancement and deployment of quantum processing units (QPUs) demands parallel development of quantum algorithms, which can leverage this evolving technology to address open scientific challenges. Many near-term quantum algorithms have been proposed, among which state preparation, energy estimation, and dynamics simulation methods are particularly pertinent to physics, chemistry, and materials science research~\cite{Bharti2022noisyiq, cerezo2021variational, bauer2020quantum, rmp_qcc,oftelieSimulatingQuantumMaterials2021, gqce, Vorwerk2022QuantumET}. Representative practical calculations on QPUs include post-quench or periodically driven nonequilibrium dynamics and correlation function measurement utilizing Trotter-decomposed circuits~\cite{kim2023scalable, kimEvidenceUtilityQuantum2023,  Trotter_dynamics_Knolle, Chen2022, chen2023problemts, Trotter_dynamics_Lawrence, mi2022time, zhangDigitalQuantumSimulation2022,freyRealizationDiscreteTime2022a,delreRobustMeasurementsPoint2024}, ground state energy estimation using auxiliary-field quantum Monte Carlo guided by trial states prepared on a QPU (QC-AFQMC)~\cite{Huggins2022Unbiasingfq}, and state preparation by optimizing parameterized quantum circuits (PQCs)~\cite{zhao2023Orbitalpc, Zhu2020GenerationOT, vqe_pea_h2}. Near-term applications with a large number of qubits favor quantum dynamics simulations with Trotter circuits owing to their modest circuit depth scaling and the possibility of matching the required gates to the hardware connectivity~\cite{Trotter_dynamics_Knolle, Chen2022, kim2023scalable, kimEvidenceUtilityQuantum2023}. The execution of variational algorithms like the variational quantum eigensolver (VQE), on the other hand, are generally more demanding~\cite{vqe_theory, alan_ucc2018, cerezo2021variational, tilly2022variational, liBenchmarkingVariationalQuantum2023}. Here, the limiting factors are that deep circuits are often needed to accurately represent the ground state, in addition to the need to perform a costly high-dimensional classical optimization of a (generally nonconvex) noisy cost function that often experiences barren plateaus~\cite{mcclean2018barren, wang2021noise}. The quantum-classical feedback loop in VQE results in large measurement overheads, although some of the difficulties can be alleviated with alternative algorithms such as the quantum imaginary time evolution~\cite{VQITE, qite_chan20, AVQITE}.

Alternatively, quantum subspace expansion (QSE) algorithms have been proposed to simulate ground and excited states as well as nonequilibrium dynamics. The quantum Krylov subspace expansion and its generalizations utilize subspaces that are generated along the trajectory of quantum imaginary-~\cite{qite_chan20, AVQITE} or real-time~\cite{MRSQK, QFD, seki2021quantumpower, VQPE, cortes2022quantumks, cortes2022fastqs} evolution from various initial states. These algorithms are still challenging to implement on noisy intermediate-scale quantum (NISQ) devices. For example, measuring off-diagonal elements of Hamiltonian and overlap matrices in the nonorthogonal subspace involves the use of the Hadamard test, which generally involves deep Trotter circuits controlled by an ancilla qubit. For practical calculations on current QPUs, QSE methods where the basis states are prepared by applying tensor products of Pauli operators on a reference state are preferable~\cite{colless2018computationms, IQAE, takeshita2020, limfastfnp}. They only require direct measurements, even though a substantial number of measurements is often needed.

Here, we focus on the iterative quantum-assisted eigensolver (IQAE)~\cite{IQAE} to highlight the 
general idea that VQE and QSE can be synergistically integrated (VQESE) to improve the accuracy of ground state energy estimation 
on noisy intermediate-scale quantum (NISQ) hardware. 
We systematically investigate the benefits and trade-offs of balancing the depth of the VQE ansatz (parametrized by the number $L$ of circuit layers) with the dimension of the QSE basis (parametrized by the expansion moment $K$) to obtain a desired ground state energy accuracy. Since we specifically explore IQAE along these two different directions parametrized by $L$ and $K$, we refer to the algorithm as the ``paired iterative quantum-assisted eigensolver'' (PIQAE) in the following. 

Our results emphasize the capacity of PIQAE to enhance the accuracy of ground state calculations by choosing variational ans\"atze of circuit depth $L$ that are compatible with hardware errors and subsequently refining the variationally optimized state via expansion in a fine-grained Krylov subspace (to be defined below) up to an expansion moment $K$ that is allowed by the classical and quantum computational budget. The expansion increases the expressivity of the resulting wavefunction without complicating trainability, since the diagonalization of the low-energy Hamiltonian in the Krylov subspace (whose coefficients are obtained via quantum measurements) is performed fully classically. Thus, one can consider various PQC forms of controlled depth, including adaptively generated, problem-specific ans\"atze~\cite{grimsleyAdaptiveVariationalAlgorithm2019, AVQITE}.
To demonstrate the versatility of PIQAE, we here employ the layered Hamiltonian Variational Ansatz (HVA), which offers flexibility in adjusting circuit depth via the layer number $L$.

Through statevector simulations of 1D and 2D transverse and mixed-field Ising models (TFIMs and MFIMs), we show that PIQAE achieves accurate ground state solutions along the pair of parameter axes defined by the VQE ansatz depth $L$ and the QSE moment $K$. Our findings reveal that the convergence with $K$ occurs more rapidly with increasing $L$.  
We then investigate the effects of hardware noises, focusing first on shot noise. Noise poses a severe challenge since the generalized eigenvalue problem (defined by noisy Hamiltonian and overlap matrices) is not bounded from below by the true ground state energy $E_G$. One therefore needs to impose a criterion to choose the optimal subspace dimension $\mathcal{M}_o$ to obtain accurate energy estimates. Here, we propose a criterion based on the trace of the overlap matrix, which we find to provide reliable and accurate energy estimates across different system sizes and noise levels. 
We then perform PIQAE calculations of $5$- and $16$-site MFIMs on the \texttt{ibmq\_quito} and \texttt{ibmq\_guadalupe} QPUs. Using quantum error mitigation techniques, we observe an order of magnitude improvement of the energy error per site compared to the initial VQE energy. Finally, we demonstrate on the noisy simulator \texttt{fake\_guadalupe} how one can approach the exact solution of the 16-site MFIM using controlled quantum error mitigation based on probabilistic error reduction, which includes noise tomography.   

\section{Method}
\label{sec:methods}
For completeness, we now describe the iterative quantum-assisted eigensolver of Ref.~\cite{IQAE}, which is a NISQ-compatible algorithm to obtain the ground state energy of a Hamiltonian based on the Krylov subspace (KS) expansion. Let us define a generic qubit Hamiltonian 
\begin{equation}
    \h=\sum_{j=1}^{N_H} c_j \hat{P}_j
\end{equation}
as a weighted sum of Pauli strings ($\hat{P}_j=\{\hat{I}, \hat{X}, \hat{Y}, \hat{Z}\}^{\otimes N}$) for an $N$-qubit system, where $\{\hat{I}, \hat{X}, \hat{Y}, \hat{Z}\}$ are the identity and Pauli operators, respectively. The coefficients $c_j$ are real owing to the fact that $\h$ is Hermitian.

Starting from a (normalized) state $\ket{\Psi}$ which can be prepared on a quantum computer, the fine-grained Krylov subspace (FGKS) of moment $K$
is defined as as the union
\begin{equation}
    \mathbb{CS}_K \equiv \bigl\{ \ket{V_j}\bigr\}_{j=1}^{N_K} = \cup_{k=0}^K \mathbb{S}_k \,,
\end{equation}
where $\mathbb{S}_k \equiv \{\ket{V_j} \equiv \hat{P}_j \ket{\Psi}\}_{\hat{P}_j \in \mathbb{P}_k}$. Here we define
\begin{equation}
    \mathbb{P}_k\equiv \mathbb{H}^k\setminus\cup_{l=0}^{k-1}\mathbb{H}^l
\end{equation}
with $\mathbb{H}^k$ being the set of Pauli strings in $\h^k$ and $\mathbb{H}^0 \equiv \{\hat{I}^{\otimes N}\}$. Thus, $\mathbb{P}_k$ is the set of Pauli strings in $\h^k$ that did not appear in any lower power of $\h^l$ ($l < k$). 
Note that $N_K$ denotes the number of terms in the overcomplete basis $\ket{V_j}$ that spans $\mathbb{CS}_K$, which may be different from the dimension of the FGKS, because some of the vectors $\ket{V_j}$ can be linearly dependent. 

An approximation to the ground state is given by the lowest energy eigenvector $\ket{\Psi_\text{KS}}$ of the generalized eigenvalue equation 
\be
\sum_{j}\mathcal{H}_{ij}V_{jk} = \sum_{j}\mathcal{S}_{ij}V_{jk}\lambda_k \,. 
\label{eq:gee}
\ee
Here, $\mathcal{H}_{ij} = \braket{V_i|\h|V_j}$ and $\mathcal{S}_{ij} = \braket{V_i|V_j}$ are the Hamiltonian and overlap matrices, respectively, 
given in terms of the overcomplete basis vectors of the subspace $\mathbb{CS}_{K}$. The $k$th eigenvalue $\lambda_k$ is associated with (column) eigenvector $V_{jk}$. Notice that one can measure the matrix elements of $\mathcal{H}$ and $\mathcal{S}$ directly on a quantum computer, as they amount to the expectation values of Pauli strings in state $\ket{\Psi}$. In the following, we use $\mathcal{D}_K = |\mathbb{CP}_K|$ to denote the size of the set of unique Pauli strings in 
\begin{equation}
    \mathbb{CP}_K\equiv \bigl\{\hat{P}_i \hat{P}_k \hat{P}_j \bigr\}_{\hat{P}_i, \hat{P}_j \in \cup_{k=0}^K \mathbb{P}_k}^{\hat{P}_k \in \cup_{k=0}^1 \mathbb{P}_k}
\end{equation}
that are used in the computation of $\mathcal{H}$ and $\mathcal{S}$. As discussed in Ref.~\cite{IQAE}, the accuracy the IQAE ground state solution depends on the initial (or zero-moment) state $\ket{\Psi}$ and the moment $K$ of the subspace expansion. The critical moment $K_c$ for convergence to a desired accuracy is problem-dependent and upper bounded by the rank of the Hamiltonian $\h$. 

Within PIQAE one approaches this dependence on $\ket{\Psi}$ and $K$ systematically with the goal of improving the ground-state solution along these two different directions, while balancing the quantum and classical computational costs according to the available hardware. The quality of the zero-moment state $\ket{\Psi}$ can be described by the depth of a state preparation circuit. For the numerical studies here, we adopt the Hamiltonian Variational Ansatz (HVA) to prepare the zero-moment state,
\be
\ket{\Psi} = \Pi_{l=1}^{L} \mathcal{U}(\bth_l)\ket{\Psi_0}, \label{eq:HVA}
\ee
where each layer of parameterized unitaries $\mathcal{U}(\bth_l)$ depends on the Hamiltonian and the circuit depth is proportional to total number of layers $L$. The reference state $\ket{\Psi_0} = H^{\otimes N} \ket{0}$, where $H$ is the Hadamard gate and $\ket{0} \equiv \ket{0}^{\otimes N}$, is taken to be a uniform product state in the $x$ basis.
 
\section{Models and HVA ans\"atze}
\label{sec:models}
We consider an $N$-site spin-$1/2$ MFIM with Hamiltonian
\begin{equation}
    \h = J\sum_{\braket{ij}} \hat{Z}_i \hat{Z}_j + \sum_i (h_x\hat{X}_i + h_z\hat{Z}_i), \label{eq:hamiltonian}
\end{equation}
where $\hat{X}$ and $\hat{Z}$ are Pauli matrices, $J$ is the nearest-neighbor coupling amplitude, and $h_x$ and $h_z$ are the transverse and longitudinal magnetic field strengths, respectively.
$\braket{ij}$ indicates that the summation is restricted to nearest-neighbors. 
The model reduces to the TFIM by setting $h_z = 0$. In the following calculations, we set $J=-1$ to be a ferromagnetic coupling and use $\abs{J}=1$ as the energy unit. The other parameters, $h_x$ and $h_z$, are given within each specific calculation below, where we consider systems in both one and two dimensions. In 2D, we consider both a square lattice geometry and the heavy-hex lattice geometry of the IBM QPUs.

For the MFIM, we adopt the following form for the one-layer unitaries of the HVA ansatz:
\be
\mathcal{U}(\bth_l) = \mathcal{U}^x(\gamma_l) \mathcal{U}^z(\beta_l) \mathcal{U}^{zz}(\alpha_l)\label{eq:ansatz},
\ee
where
\begin{align}
\mathcal{U}^{zz}(\alpha) &= \exp \left(-i\frac{\alpha}{2} \sum_{\left\langle ij \right\rangle} Z_i Z_j \right), \label{eq: zz}\\ 
\mathcal{U}^{z}(\beta) &= \exp \left(-i\frac{\beta}{2} \sum_i Z_i \right), \\
\mathcal{U}^{x}(\gamma) &= \exp \left(-i\frac{\gamma}{2} \sum_i X_i \right).
\end{align}
The unitaries above originate from the exchange coupling term and the longitudinal and transverse field terms in the Hamiltonian, respectively. The associated variational parameters are denoted by $\alpha,\;\beta$, and $\gamma$. For the TFIM, the one-layer unitaries take a simpler form since the longitudinal field term vanishes:
\be
\mathcal{U}(\bth_l) = \mathcal{U}^x(\gamma_l) \mathcal{U}^{zz}(\alpha_l).
\ee

\begin{figure}
    \centering
    \includegraphics[width=\linewidth]{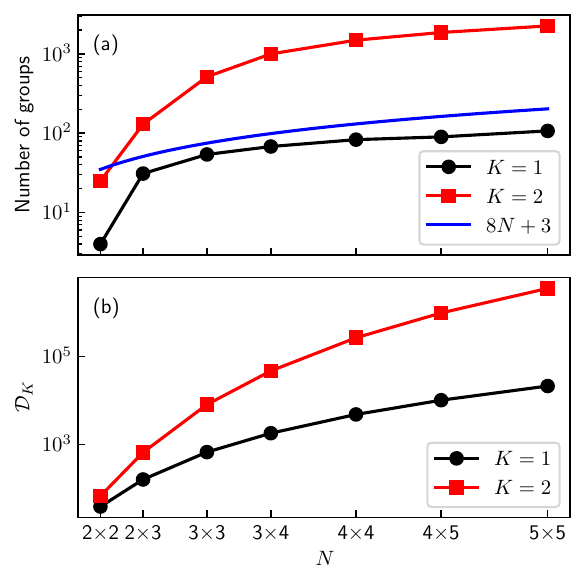}
    \caption{\textbf{System-size dependence of the number of groups of commuting Pauli strings in $\mathbb{CP}_K$ for the 2D TFIM.} (a) Number of self-commuting groups $N_\text{g}$ of Pauli strings as a function of system size $N$ for moments $K=1$ (black circles) and $K=2$ (red squares) for the 2D TFIM. These numbers are obtained using a heuristic greedy coloring algorithm with vertices of largest degree first~\cite{kubale2004graph}. The blue line represents an analytical upper bound for $K=1$. (b) Total number of Pauli strings $\D_k$ in $\mathbb{CP}_K$ for PIQAE calculations of the 2D TFIM on the square lattice with $K=1$ (black circles) and $K=2$ (red squares).
    }
    \label{fig:scaling}
\end{figure}

\section{Quantum resource estimation}
The quantum resources required for a PIQAE calculation comprise the PQCs for preparing the zero-moment state $\ket{\Psi}$ in Eq.~\eqref{eq:ansatz} and measurement circuits for the $\D_K$ Pauli strings in $\mathbb{CP}_K$. Since the preparation of the zero-moment state is a typical VQE calculation with a particular choice of ansatz, for which the associated quantum resource requirements have been extensively discussed~\cite{vqe_theory, alan_ucc2018, tilly2022variational}, we focus on the contribution from measurements needed for the QSE. Here the quantum resource cost is controlled by the number of groups of commuting Pauli strings in $\mathbb{CP}_K$, such that all Pauli strings in a given group can be measured simultaneously. In Fig.~\figref{fig:scaling}{(a)}, we plot the number of groups of Pauli strings $N_\text{g}$ as a function of system size $N$ for the 2D TFIM. As the partitioning of observables into self-commuting groups is equivalent to an NP-hard graph coloring problem, we use a heuristic greedy coloring algorithm which includes vertices of largest degree first~\cite{kubale2004graph} to determine $N_\text{g}$ for two different values of the expansion moment $K$. As expected, $N_\text{g}$ for $K=2$ (red squares) is much larger than that for $K=1$ (black circles) due to the larger $\D_K$, as shown in Fig.~\figref{fig:scaling}{(b)}. Nevertheless, $N_\text{g}$ shows a modest, approximately linear growth for large $N$. In practice, this places within reach system sizes $N$ on the order of a hundred, which can be challenging to simulate classically. 

For reference, one can get a loose upper bound for $N_\text{g}$ by considering the specific case of the TFIM [Eq.~\eqref{eq:hamiltonian}]. For moment $K=1$, $\mathbb{CP}_K$ can be split into the following self-commuting groups: all $Z$ site-wise, all $X$, all $Y$, all $Z$ except $X_i$, all $Z$ except $Y_i$, all $X$ except a nearest neighbor pair $ZZ$, all $X$ except a nearest neighbor pair $YZ$, and all $X$ except a nearest neighbor pair $ZY$, which amounts to the upper bound
\be
N_\text{g}\leq1+1+1+N+N+2N+2N+2N=8N+3\,, \label{eq: ub}
\ee
denoted by the blue line in Fig.~\figref{fig:scaling}{(a)}. Notice that this analytical upper bound is close to the greedy coloring results and that the overestimation becomes larger as $N$ increases.

\begin{figure*}
    \centering
    \includegraphics[width=\linewidth]{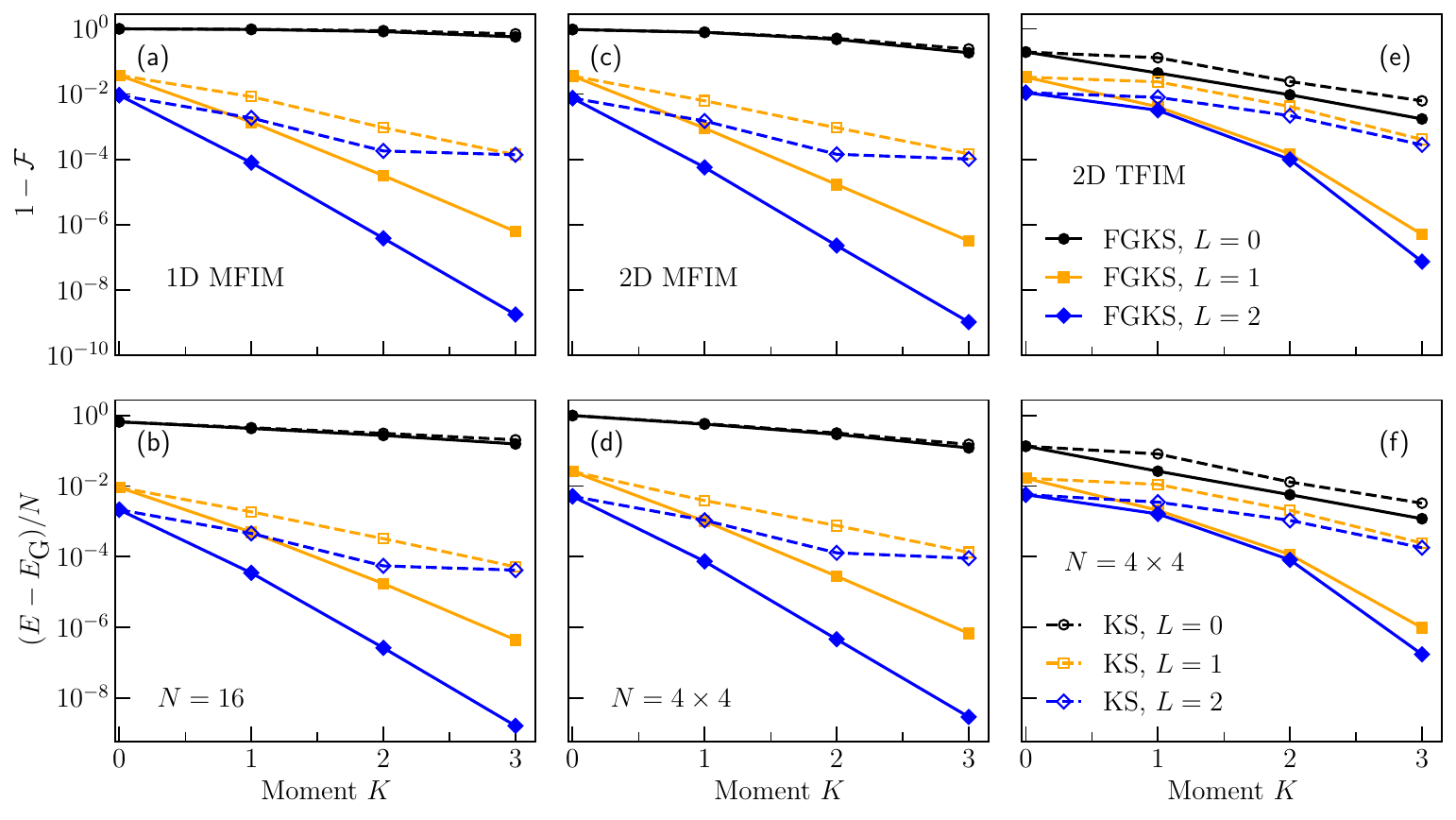}
    \caption{\textbf{Fidelity and energy convergence of PIQAE calculations with HVA ansatz layer number $L$ and FGKS expansion moment $K$ using statevector simulator.} (a) State infidelity, $1-\mathcal{F} \equiv 1- \abs{\ov{g(K,L)}{G}}^2$, as a function of $K$ with $L=0$ (black filled circles), $L=1$ (orange filled squares), and $L=2$ (blue filled diamonds) for the 1D MFIM of $N=16$ sites. Here $\ket{g(K,L})$ is the FGKS ground state, and $\ket{G}$ is the ED ground state. (c, e) Similar to (a) but for the 2D MFIM and TFIM on an $N=4\times4$ square lattice. (b) Similar to (a) but for the energy error per site $\varepsilon=[E_g(K,L)-E_\text{G}]/N$, with $E_g(K,L)\equiv \Av{g(K,L}{\h}$ and $E_G\equiv \Av{G}{\h}$. (d, f) Similar to (b) but for 2D MFIM and TFIM with $N=4\times4$ sites. For reference, the state infidelity and energy error from the same type of subspace expansion calculations but within the conventional Krylov subspace (KS), $\{\h^k \ket{\Psi}\}_{k=0}^K$, are also presented in each panel, with $L=0$ (black open circles), $L=1$ (orange open squares), and $L=2$ (blue open diamonds). (See text for Hamiltonian parameters.)
    }
    \label{fig:N16}
\end{figure*}

\section{Statevector simulations}
In this section, we demonstrate numerically the performance of the PIQAE method in ground state calculations by investigating its dependence on the HVA ansatz layer number $L$ and the FGKS expansion moment $K$. We choose here the 1D $N=16$ MFIM, and the 2D $N=4\times 4$ square lattice TFIM and MFIM as examples, and perform statevector simulations. The transverse field strength is set to $h_x= -3.05 $ for both the 2D TFIM and MFIM, thus placing the former in the vicinity of a ferromagnetic-paramagnetic critical point~\cite{bloteClusterMonteCarlo2002}. On the other hand, for the 1D MFIM we set $h_z = - h_x/2 = 0.5$, while $h_z = -h_x/2 = 1.525$ for the 2D MFIM. In the calculations, we vary $L$ from zero (i.e., $\ket{\Psi}=\ket{+}^{\otimes N}$ is a product state) to $L=2$, and $K$ from zero (i.e., with a single state $\ket{\Psi}$ in the FGKS) to $K=3$. 

In order to evaluate the accuracy of PIQAE calculations, we consider in Figure~\ref{fig:N16} as a figure of merit the energy error per site 
\begin{equation}
    \varepsilon = \frac{E - E_G}{N} \,,
\end{equation}
where $E_G$ is the exact ground state energy found via exact diagonalization (ED) of the Hamiltonian in Eq.~\eqref{eq:hamiltonian}, and the infidelity 
\begin{equation}
    1-\mathcal{F} = 1 - |\braket{g(K,L)|G}|^2
\end{equation}
of the FGKS ground state $\ket{g(K,L)}$ with respect to the ED state $\ket{G}$. 

Generally, the fidelity and energy improve when increasing the number of ansatz layers $L$ or the expansion moment $K$, as expected due to enhanced expressivity of the wavefunction in the FGKS.
Importantly, the convergence rate with $K$ clearly becomes faster with increasing $L$. When $L=0$, where the zero-moment state $\ket{\Psi}$ is reduced to a simple product state $\ket{+}^{\otimes N}$, the infidelity and energy error reduce by about one order of magnitude when $K$ increases from zero to three for the 2D MFIM. In contrast, with $L=2$, they reduce by about six orders of magnitude in the same range of $K$ for the 2D MFIM. One can reach an accuracy of $1-\mathcal{F}\approx 10^{-4}$ and $\varepsilon \approx 10^{-4}$ (relative error $\sim 0.01\%$) from the PIQAE calculations of the MFIM either for $L=1$ and $K=2$ or for $L=2$ and $K=1$. Note that the convergence behaviors of the 1D and 2D MFIMs shown in Fig.~\ref{fig:N16}~\hyperref[fig:N16]{(a-d)} resemble each other, which shows that the dimensionality does not play an important role in the performance of the algorithm. 

Compared with the MFIM results, the PIQAE calculation for 2D TFIM shows overall slower convergence behavior of the state infidelity and energy error with $L$ and $K$, which is consistent with the model being close to quantum criticality. Here, the convergence rate with $K$ is similar for $L=1$ and $L=2$, but notably faster than for $L=0$. Still, both calculations with $L=1,2$ reach an accuracy of $1-\mathcal{F}\approx 10^{-4}$ and $\varepsilon\approx 10^{-4}$ at $K=2$. For reference, in Fig.~\ref{fig:N16} we also plot the results from simulations similar to PIQAE, but calculated within the conventional Krylov subspace (KS) spanned by the set of vectors $\{\h^k \ket{\Psi}\}_{k=0}^K$, rather than the FGKS. Generally, the KS expansion results show a slower convergence behavior than PIQAE calculations. For example, the state infidelity is about $9\times 10^{-4}$, and energy error per site is about $7\times 10^{-4}$ for the KS results of 2D TFIM, compared to $1-\mathcal{F}\approx 2\times 10^{-5}$ and $\varepsilon\approx 2\times 10^{-5}$ for the PIQAE results at $L=1$ and $K=2$.

This demonstrates the flexibility of ground state preparation using PIQAE, where the simulation strategy can be tailored to quantum hardware with specific error rates. The classical resource cost set by by $K$ (which fixes the size $\mathcal{D}_K$ of $\mathcal{H}_{ij}$ and $\mathcal{S}_{ij}$) and the quantum resource cost tied to $L$ for the circuit depth and $K$ for the number of measurement bases needed, $N_\text{g}$, can be tuned to reach optimal results.

Finally, we make a brief technical note. In the above PIQAE calculations, the ground state is obtained by numerically diagonalizing the generalized eigenvalue equation~\eqref{eq:gee}. In practice, the Hamiltonian matrix $\mathcal{H}$ and overlap matrix $\mathcal{S}$ may not be full rank due to a linear dependence of some the FGKS vectors in $\mathbb{CS}_K$. This issue is resolved by the Hamiltonian regularization procedure, which reconstructs a Hamiltonian matrix in a smaller subspace spanned by the set of eigenvectors of $\mathcal{S}$ with eigenvalues larger than a threshold, which we set to $\xi_\text{c} = 10^{-6}$. We refer to the dimension of this truncated subspace as $\mathcal{M}$. We find that Hamiltonian regularization can reduce the corresponding matrix dimension, with an increasing reduction rate as a function of expansion moment $K$, ranging from approximately 33\% for $K=1$ and 75\% for $K=3$ in the calculations shown in Fig.~\ref{fig:N16}.

\begin{figure}[t!]
    \centering
    \includegraphics[width=\linewidth]{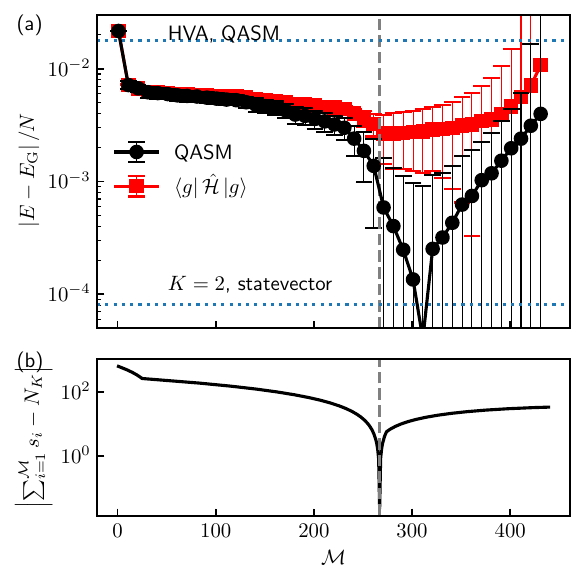}
    \caption{\textbf{QASM simulator results for $4\times 4$ TFIM} (a) Energy error per site $\varepsilon = |\Tilde{E}_g-E_G|/N$ of the estimated FGKS ground state $\ket{g}$ in the presence of shot noise as a function of the truncated subspace dimension $\mathcal{M}$. The zeroth moment state is an HVA ansatz with $L=1$ layer, the expansion moment is $K=2$, and we used $M_\text{s} = 2^{14}$ measurement shots for each circuit. 
    For comparison, we also show the error $|\avihi{g}{\h} - E_G|/N$ of the exact energy of the state $\ket{g}$ as a function of $\mathcal{M}$ (red squares). This quantity is a figure of merit of the ground state $\ket{g}$. 
    (b) Absolute difference $ |\sum_{i=1}^{\mathcal{M}} s_i - N_K|$ between the 
    sum of the largest $\mathcal{M}$ eigenvalues $s_i$ of $\Tilde{\mathcal{S}}$ and the trace of the overlap matrix $\text{Tr} \, \mathcal{S} = N_K$ as a function of $\mathcal{M}$. The dashed grey line indicates $\mathcal{M}_\text{o}$.
    Error bars in (a) are standard deviations estimated based on $10$ sets of $\Tilde{\mathcal{H}}$ and $\Tilde{\mathcal{S}}$, obtained from sampling the expectation values of Pauli strings in $\mathbb{CP}_K$ according to the multivariate normal distribution specified by the measured means and standard errors.
    For clarity, the results are shown at every $10$th point.
    We set $h_x= -3.05$ for the 2D TFIM as in Fig.~\ref{fig:N16}. 
    }
    \label{fig:qasm}
\end{figure}

\section{Shot-noise effects}
The statevector simulations assume infinite precision for the expectation value $\braket{\hat{O}}$ of an observable $\hat{O}$, while in practice $\braket{\hat{O}}$ is always subject to statistical errors due to the finite number of samples (or shots) $M_\text{s}$ for each measurement, even within a fault-tolerant quantum computer. Therefore, it is crucial to assess and mitigate the impact of shot noise on the PIQAE calculations. Specifically, we are interested to determine how statistical noise in the Hamiltonian $\Tilde{\mathcal{H}}$ and overlap matrix $\Tilde{\mathcal{S}}$ impact the accuracy of the lowest eigenvalue $E_{g}(K,L)$ of Eq.~\eqref{eq:gee}. This issue 
has been discussed previously in the QSE literature~\cite{VQPE}, and one possible solution is to use the Hamiltonian regularization method adopted in the statevector simulations but with the truncated subspace dimension $\mathcal{M}$ fixed by a modified eigenvalue size threshold $\xi_\text{c}$ tied to the number of shots $M_\text{s}$. 
Here, however, we propose using a different approach based on preserving the trace of the overlap matrix, $\text{Tr} \mathcal{S}$. Since the vectors $\{\ket{V_j} \}$ are normalized, we find $\text{Tr} \mathcal{S} = N_K$ in the noiseless case. After diagonalizing $\mathcal{S}$ (with eigenvalues $s_i$ ordered in decreasing size), we impose the criterion $\sum_{i=1}^{\mathcal{M}_o} s_i = N_K$ as a way to select the optimal number $\mathcal M = \mathcal{M}_o$ of states to keep. 

Figure~\ref{fig:qasm} shows results for the 2D square lattice TFIM with $N=4\times 4$ sites obtained on the quantum assembly language (QASM)-based simulator as implemented in Qiskit~\cite{Qiskit}. We set $L=1$ for the HVA ansatz and $K=2$ for FGKS expansion moment. Since the focus here is on the impact of shot noise on the solution of the generalized eigenvalue equation [Eq.~\eqref{eq:gee}], we fix the variational parameters of the HVA ansatz to those obtained from statevector simulations ($\alpha = 0.154, \gamma = 0.785$ ). We choose $M_\text{s} = 2^{14}$ for the number of measurements of each Pauli string in $\mathbb{CP}_K$. Figure~\ref{fig:qasm}~\hyperref[fig:qasm]{(a)} plots the energy error per site $\varepsilon = |\tilde{E}_g - E_G|/N$ as a function of truncated subspace dimension $\mathcal{M}$. Here, $\tilde{E}_g \equiv \Tilde{E}_g(K,L,\mathcal{M})$ is the lowest energy of the measured (noisy) Hamiltonian matrix $\Tilde{\mathcal{H}}_{i j} = \mel{s_i}{\Tilde{\mathcal{H}}}{s_j}$ in the subspace spanned by the eigenvectors $\{\ket{s_j}\}$ of the noisy $\Tilde{\mathcal{S}}$ with the largest $\mathcal{M}$ eigenvalues. From now on we shall use $\Tilde{O}$ to emphasize the observable $O$ evaluated in the presence of shot noise or device errors where needed to avoid confusion. The associated ground state $\ket{g} \equiv \ket{g(K,L,\mathcal{M})}$ reads
\be
\ket{g} = \sum_{i=1}^{\mathcal{M}} c_i \ket{s_i} = \sum_{k=1}^{N_K} c'_k \ket{V_k}, \label{eq:em}
\ee
where $c_i$ and $c'_i$ are expansion coefficients.
The energy error per site starts at $\varepsilon(\mathcal{M}=1) = 0.0217(2)$, which is slightly higher than that of the HVA energy $(E_{\text{HVA}}-E_G)/N = 0.017(3)$ (upper dotted line in Fig.~\figref{fig:qasm}{(a)}). Note that the basis vector of the truncated subspace with $\mathcal{M}=1$ does not necessarily coincide with the zeroth-moment HVA starting state $\ket{\Psi}$ owing to the subspace construction method involving the diagonalization of $\mathcal{S}$. The energy error initially decreases with increasing $\mathcal{M}$, and reaches a minimum of about $\varepsilon(\mathcal{M} = 310) = 0.1(9) \times 10^{-4}$, which is comparable to the $K=2$, $L=1$ statevector result $\varepsilon_{\text{SV}} = 0.8\times 10^{-4}$, denoted by the lower dotted line in Fig.~\figref{fig:qasm}{(a)}. 

The above discussion assumes knowledge of the exact reference point $E_G$, for which the minimal error is obtained. Since this is unknown for larger systems, one needs a criterion independent of $E_G$ to determine the optimal value of $\mathcal M = \mathcal{M}_\text{o}$ for noisy calculations. The $\mathcal{S}$-trace criterion takes note that $\text{Tr} \mathcal{S} = N_K$ since the $N_K$ vectors $\{\ket{V_j}\}$ spanning the FGKS are normalized (recall that this provides an overcomplete basis for the FGKS). Therefore, the cumulative sum of the eigenvalues $\sum_{i=1}^{\mathcal{M}} s_i$ of the overlap matrix $\mathcal{S}$ evaluated without noise is upper bounded by the matrix dimension $N_K$. Note that the eigenvalues are ordered such that $s_i \geq s_{i+1}$ for all $i$. While this does not hold in the presence of noise, we propose the $\mathcal{S}$-trace criterion to determine $\mathcal{M}_\text{o}$ by minimizing the distance 
\begin{equation}
    \mathcal{M}_\text{o} = \min_{\mathcal{M}} \abs{\sum_{i=1}^{\mathcal{M}} s_i - N_K} \,.
\end{equation}
In Fig.~\figref{fig:qasm}{(b)} we plot the distance as a function of $\mathcal{M}$. Following the $\mathcal{S}$-trace criterion, we determine $\mathcal{M}_\text{o} = 267$. This gives $\varepsilon = (1 \pm 1)\times 10^{-3}$, which is close to the ranges of estimation above ($0.1(9) \times 10^{-4}$). We include these results in Table~\ref{tab:results}, which summarizes the accuracy of the ground state energy obtained for the different noisy simulations performed in this work. As a figure of merit for the estimated ground state $\ket{g}$ in Eq.~\eqref{eq:em}, we show its exact energy expectation value, $E_g=\avihi{g}{\h}$ (after normalizing $\ket{g}$), as a function of $\mathcal{M}$ in red squares in Fig.\figref{fig:qasm}{(a)}. Note that we use the exact Hamiltonian from Eq.~\eqref{eq:hamiltonian} here. One can see that the two curves in Fig.\figref{fig:qasm}{(a)} exhibit a similar behavior, showing a larger deviation around $\mathcal{M} \gtrsim 200$, and starting a slight upturn around $\mathcal{M}_\text{o} = 267$, above which they continue to increase. The presence of a minimum of $E_g=\avihi{g}{\h}$ at $\mathcal{M}_\text{o}$ validates the $\mathcal{S}$-trace criterion and provides a sense of its operational meaning.

\begin{table}[t]
    \centering
    \begin{tabular}{*{6}{c}} 
     \toprule
     Model & Backend & $K$& $L$ &$\varepsilon(\mathcal{M}_\text{o})$ 
     & $\varepsilon_{\text{rel}}(\mathcal{M}_\text{o})$
     \\
     \midrule
     $4\times 4$ TFIM & QASM & $2$ & $1$ & $1(1) \times 10^{-3}$ & 0.03(3)\% \\
     $5$-site MFIM & \texttt{ibm\_quito} & $2$ & $2$ & $2.9(6) \times 10^{-3}$ & 0.18(4)\% \\
     $16$-site MFIM & \texttt{ibm\_guadalupe} & $1$ & $1$ & $0.05(4)$ & 3(2)\% \\
     $16$-site MFIM & \texttt{fake\_guadalupe} & $1$ & $1$ &  $1(1) \times 10^{-3}$ & 0.06(6)\% \\
     \bottomrule
    \end{tabular}
    \caption{Summary of ground state energy accuracies obtained in the different simulations in the presence of noise. Here, $\varepsilon(\mathcal{M}_\text{o}) = |\tilde{E}_g(K,L,\mathcal{M}_\text{o}) - E_G|/N$ denotes the error per site at the optimal value of $\mathcal{M}_\text{o}$
    and $\varepsilon_\text{rel}(\mathcal{M}_\text{o}) = |\tilde{E}_g(K,L,\mathcal{M}_\text{o}) - E_G|/E_G$ the relative error.
    }
    \label{tab:results}
\end{table}

\begin{figure}[t!]
    \centering
    \includegraphics[width=\linewidth]{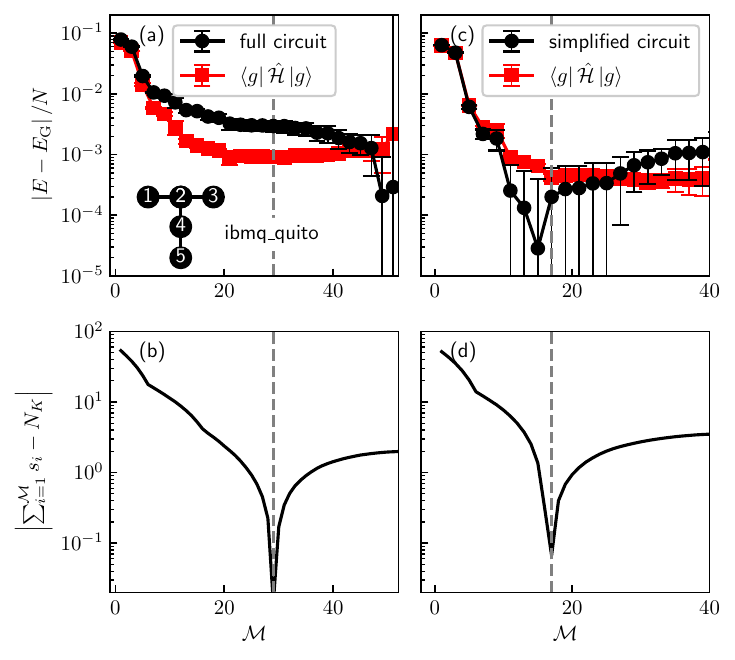}
    \caption{\textbf{PIQAE results on \texttt{ibmq\_quito} QPU.} (a) Energy error per site $\varepsilon= |\Tilde{E}_g-E_\text{G}|/N$ (black dots) as a function of subspace dimension $\mathcal{M}$ for a 5-site MFIM on the \texttt{ibmq\_quito} lattice (see inset). For comparison, we show the error $|\avihi{g}{\h}-E_G|/N$ of the exact energy of the FGKS ground state $\ket{g}$ obtained from the noisy $\Tilde{\mathcal{H}}_{ij}$ and $\Tilde{\mathcal{S}}_{ij}$ (red squares). 
    (b) Absolute difference $ |\sum_{i=1}^{\mathcal{M}} s_i - N_K|$ between the sum of (sorted) eigenvalues $s_i$ and the dimension of the noiseless overlap matrix $\text{Tr} \, \mathcal{S} = N_K$ as a function of $\mathcal{M}$.
    Panels (c,d) show the same information as (a,b) except that the entangling gates have been omitted from the HVA ansatz owing to the small optimal angle $\alpha<10^{-3}$. The error bars in (a,c) are obtained in the same way as in Fig.~\ref{fig:qasm}.
    We set $h_x= -1$ and $h_z = 0.5$, and use $M_s = 2^{14}$ shots to measure each group of commuting Pauli strings in $\mathbb{CP}_K$.
    }
    \label{fig:qpu5}
\end{figure}

\section{Quantum hardware calculations}
\label{qpu}
\subsection{$5$-site MFIM simulations on \texttt{ibm\_quito}}
To demonstrate PIQAE calculations on quantum hardware, we first choose a MFIM of $N=5$ on a lattice matching the qubit layout of the \texttt{ibmq\_quito} QPU, as shown in Fig.~\ref{fig:qpu5}. An HVA ansatz of $L=1$ is adopted, and the FGKS moment is set to $K=2$. Similar to the benchmark calculations with shot noise, we optimize the parameters of HVA using the statevector simulator to focus on the impact of device errors on the subspace expansion step. We obtain the optimal values $\alpha = -1\times10^{-8}$, $\beta = -1.09$, and $\gamma = 1.57$, which define the zeroth moment state $\ket{\Psi}$.

To obtain $\Tilde{\mathcal{H}}_{ij}$ and $\Tilde{\mathcal{S}}_{ij}$, we have to execute $N_g = 142$ measurement circuits, which yields the expectation values of $\mathcal{D}_{K=2} = 822$ Pauli operators in $\mathbb{CP}_{K=2}$. We adopt the model-free twirled readout error extinction (TREX) technique in all the QPU calculations~\cite{TREX}. Diagonalizing the generalized eigenvalue problem~\eqref{eq:gee} then yields $\tilde{E}_g(K=2, L=1, \mathcal{M})$.  

Figure~\figref{fig:qpu5}{(a)} shows that the average energy error per site, $\varepsilon(\mathcal{M})$, of the estimated ground state $\ket{g}$ decreases with $\mathcal{M}$ from its initial value $\varepsilon(\mathcal{M} =1) = 0.0788(5)$ and reaches a minimum around $\mathcal{M}=49$. However, if we were working in a regime where the exact ground state energy was not available, we would not be able to select $\mathcal M$ a posteriori according to the criterion of minimal energy error. In this case, it is desirable to implement a heuristic like the $\mathcal S$-trace criterion to select $\mathcal M$. As shown in Fig.~\figref{fig:qpu5}{(b)}, the $\mathcal{S}$-trace criterion suggests an optimal $\mathcal{M}_\text{o} = 29$, where the average error is $\varepsilon(\mathcal{M}_\text{o})=0.0029(6)$. This is more than an order of magnitude improvement compared to the initial error and also to the $L=1$ HVA energy error per site $0.13(2)$. We observe that the exact energy $E_g = \braket{g|\h|g}$ of the state $\ket{g}$ reaches a minimal error around $\mathcal{M}_\text{o} = 29$, again validating the $\mathcal{S}$-trace criterion. The minimal error of the exact energy at $\mathcal{M}_\text{o}$ is $|\braket{g|\h|g} - E_G|/N = 0.0009(1)$, which is about three times smaller than the noisy error $\varepsilon(\mathcal{M}_\text{o})$. This arises from the difference between $\h$ and the measured $\Tilde{\mathcal{H}}_{ij}$ and suggests sizable device errors. 

In the above calculation, the HVA circuit is directly transpiled to the native gates of \texttt{ibmq\_quito} without optimization. As a result, the transpiled circuit has 8 CNOT gates, which is expected to dominate the errors in the QPU calculations. To verify this conjecture, we leverage the fact that the optimal angle for the entangling gates is negligible, $\alpha \approx 10^{-8}$, and repeat the calculations with the two qubit unitaries $\mathcal{U}^{zz}$ [Eq.~\eqref{eq: zz}] removed from the HVA ansatz. Figure~\figref{fig:qpu5}{(c)} shows that the agreement between the noisy energy $\Tilde{E}_g$ and the exact $E_g = \braket{g|\h|g}$ is much better. At the optimal $\mathcal{M}_\text{o} = 17$, we find $\varepsilon(\mathcal{M}_\text{o}) = 2(4)\times 10^{-4}$ in close agreement with $|\braket{g|\h|g} - E_G|/N = 4(1) \times 10^{-4}$. 
The error is lower than in Fig.~\figref{fig:qpu5}{(a)} due to the simplified circuit free of entangling gates, which also improves the noisy overlap matrix $\Tilde{S}_{ij}$. 

\subsection{$16$-site MFIM simulations on \texttt{ibm\_guadalupe}}
Next we perform PIQAE calculations of a $16$-site MFIM on the \texttt{ibmq\_quadalupe} QPU, where we choose the spin model lattice to match the heavy-hex lattice qubit layout of the hardware. We use an $L=1$ HVA ansatz at optimal angles $\alpha = 5\times 10^{-9}$, $\beta = -1.17$, and $\gamma = 1.57$, obtained using statevector simulations. Note that rotation gates with negligible angles are also explicitly included in the QPU calculations to account for noise present in the HVA circuit at generic angles, and also to test the effect of error mitigation techniques. We set the subspace moment $K=1$ for the FGKS expansion, which amounts to measuring $\D_K = 14672$ Pauli strings in $\mathbb{CP}_K$ on the QPU to obtain $\Tilde{\mathcal{H}}_{ij}$ and $\Tilde{\mathcal{S}}_{ij}$. Grouping into qubit-wise commuting sets of Pauli strings using a greedy coloring algorithm~\cite{kubale2004graph} splits $\mathbb{CP}_{K=1}$ into $N_g = 83$ groups. We thus need to execute $N_g$ circuits with different measurement bases to obtain expectation values of all $\D_K$ Pauli operators. 
Besides adopting the TREX technique for readout calibrations, we also apply dynamical decoupling and Pauli twirling for CNOT gates to mitigate device errors. Specifically, we create $32$ equivalent circuits with dynamical decoupling and Pauli twirling, and use $2^{14}$ shots for each circuit to measure the observables. 
\begin{figure}[t!]
    \centering
    \includegraphics[width=\linewidth]{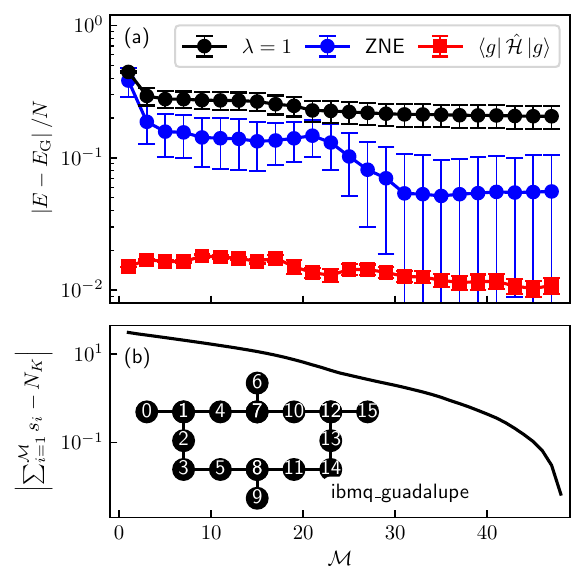}
    \caption{\textbf{PIQAE results on \texttt{ibmq\_guadalupe} QPU.} (a) Energy error per site $\varepsilon = \abs{E_g-E_\text{G}}/N$ (black dots) as a function of truncated subspace dimension $\mathcal{M}$ for a 16-site MFIM on the \texttt{ibmq\_guadalupe} heavy-hex lattice (see inset). For comparison, we show both the zero-noise extrapolated curve (blue dots) and the error $|\avihi{g}{\h}-E_G|/N$ (red squares) of the exact energy of $\ket{g}$. 
    (b) Absolute difference $|\sum_{i=1}^{\mathcal{M}} s_i - N_K|$ of the (sorted) eigenvalues $s_i$ of the noisy overlap matrix $\Tilde{\mathcal{S}}_{ij}$ and the trace of the noiseless one $N_K$ as a function of $\mathcal{M}$.
    We set $h_x= -1$ and $h_z = 0.5$ for the model and use an $L=1$ HVA ansatz and expansion moment $K=1$.
    The error bars in (a) represent the standard deviations estimated using $8$ sets of $\Tilde{\mathcal{H}}$ and $\Tilde{\mathcal{S}}$, each of which are derived from 7 out of 8 equal partitions of the full set of measurement outcomes.
    For clarity, the results are shown for every second point.
    }
    \label{fig:qpu16}
\end{figure}

As shown in Fig.~\figref{fig:qpu16}{(a)}, the average error $\varepsilon$ (black dots) rapidly drops from $\varepsilon(\mathcal{M} = 1) = 0.45(1)$ to $\varepsilon(\mathcal{M} = 3) = 0.29(5)$ in the initial two steps. This is followed by a rather slow decrease with increasing truncated subspace dimension $\mathcal{M}$ to $\varepsilon(\mathcal{M}_\text{o}) = 0.21(4)$, where $\mathcal{M}_\text{o} = 48 = \mathcal{D}_{K=1} - 1$ (see Fig.~\figref{fig:qpu16}{(b)}). The final point at $\mathcal M = \mathcal{D}_{K=1}=49$ is excluded from the analysis due to a sudden large drop in energy by over $5.5$. For reference, the $L=1$ HVA energy error per site is $0.53(1)$. In contrast, the exact energy $E_g = \braket{g|\h|g}$ of the state $\ket{g}$ experiences an error that is about 10 times smaller (red squares), and reduces only marginally from $0.015$ to a final value of $|\braket{g|\h|g} - E_G|/N = 0.011$ at $\mathcal{M}_\text{o}$. The error $\varepsilon$ for the $N=16$ model is much larger than for the $N=5$ model, which is due to the larger number of entangling gates, which is $N_{\text{CX}}= 32$ for the $N=16$ site model compared to $N_{\text{CX}} = 8$ for five sites.

To improve the energy estimation for the state $\ket{g}$, we apply the digital zero-noise extrapolation (ZNE) technique~\cite{ZNE_Temme, LaRose2022mitiqsoftware}, where we assume a uniform CNOT gate error rate of $0.98$ and perfect single-qubit gates. A noise scaling factor is varied in the range $\lambda\in [1, 1.25, 1.5, 1.75, 2]$, with $\lambda=1$ meaning no noise amplification. In addition, we construct $32$ circuits with random gate folding at each $\lambda$ using the open-source software package Mitiq~\cite{LaRose2022mitiqsoftware}, followed by application of dynamical decoupling and Pauli twirling. In Fig.~\figref{fig:qpu16}{(a)} we plot the average error of the improved estimation of $E_g$ based on ZNE with second-order polynomial fitting. ZNE generally improves the energy estimation, reducing the initial error at $\mathcal{M} = 1$ from $\varepsilon_{\lambda=1} = 0.45(1)$ to $\varepsilon_{\text{ZNE}} = 0.38(1)$ and the final one at $\mathcal{M}_\text{o}$ from $\varepsilon_{\lambda = 1} = 0.21(4)$ to $\varepsilon_{\text{ZNE}}(\mathcal{M}_\text{o}) = 0.05(4)$. The extrapolation becomes more effective at large $\mathcal{M}$, yielding a smaller error $\varepsilon$ as the subspace expands. 

We perform ZNE in the following way: first we determine the ground state $\ket{g_{\lambda=1}}=\sum_{i=1}^{N_K} c_i\ket{V_i}$ according to Eq.~\eqref{eq:em} from solving the generalized eigenvalue problem at $\lambda = 1$, i.e. with $\tilde{\mathcal{H}}_{ij}$ and $|\Tilde{\mathcal{S}}_{ij}$ obtained at $\lambda = 1$. Then, we evaluate the energy of $\ket{g_{\lambda=1}}$ with respect to the noisy Hamiltonians at other values of $\lambda > 1$ by evaluating $\sum_{ij} c_i c_j \Tilde{\mathcal{H}}_{ij}/\sum_{kl} c_k c_l \Tilde{\mathcal{S}}_{kl}$ with $\Tilde{\mathcal{H}}_{ij}$ and $\Tilde{\mathcal{S}}_{ij}$ measured at $\lambda>1$. Note that it is important to renormalize $\ket{g}$ at each $\lambda$ using the associated noisy overlap matrix.

\section{Probabilistic Error Reduction on noisy simulator}
\label{sec:PER}
In the previous section, we observed significant errors in the $L=16$ model [see Fig.~\figref{fig:qpu16}{(a)}] on the \texttt{ibm\_guadalupe} QPU, attributed to the increase in the number of entangling gates with system size. Despite applying quantum error mitigation protocols such as ZNE, dynamical decoupling, and Pauli twirling, we noted a limited improvement in energy estimates. Consequently, a more refined error mitigation strategy is essential, necessitating the learning of error channels through tomography. In this section, we use a robust and controlled quantum error mitigation protocol involving noise tailoring where initially, noise characterization is performed using Pauli noise tomography (PNT)~\cite{vandenbergProbabilisticErrorCancellation2023, flammia2020efficient}. Subsequently, error mitigation is carried out using the probabilistic error reduction (PER)~\cite{mari2021extendingqp,McDonough-AutomatedPER-2022} technique, which combines quasiprobabilistic sampling similar to probabilistic error cancellation (PEC)~\cite{ZNE_Temme} with ZNE. The PNT technique entails converting arbitrary noise channels into Pauli error channels by applying Pauli twirling to the entangling gates within each quantum circuit layer. The resulting twirled noise channel is then modeled using a sparse Lindblad model that only includes single-site and two-site Pauli operators on physically connected qubits. This is based on the assumption that strong noise correlation exists mostly between physically connected qubits, particularly nearest-neighbor pairs. This characteristic enables PNT to achieve constant scaling in the number of qubits, rendering it efficient for larger systems. Leveraging the obtained sparse Pauli noise model, we efficiently sample PER circuits from a partially inverted noise channel, where the noise strength is controlled by an external parameter $\lambda$~\cite{mari2021extendingqp,McDonough-AutomatedPER-2022}. Below, we present the resources utilized and the results of the calculations for the PNT and PER analyses performed using a completely automated open-source software called AutomatedPERTools~\cite{McDonough-AutomatedPER-2022, benjamin_mcdonough_2022_7197234}. The objective of this section is to conduct the PIQAE calculation utilizing the error-mitigated expectation values of each of the $N_g=83$ Pauli groups of commuting operators, which construct the $K=1$ FGKS on the \texttt{fake\_guadalupe} backend for the $N=16$-site MFIM, as described in Sec.~\ref{qpu}. 

\begin{figure}[t!]
    \centering
    \includegraphics[width=\linewidth]{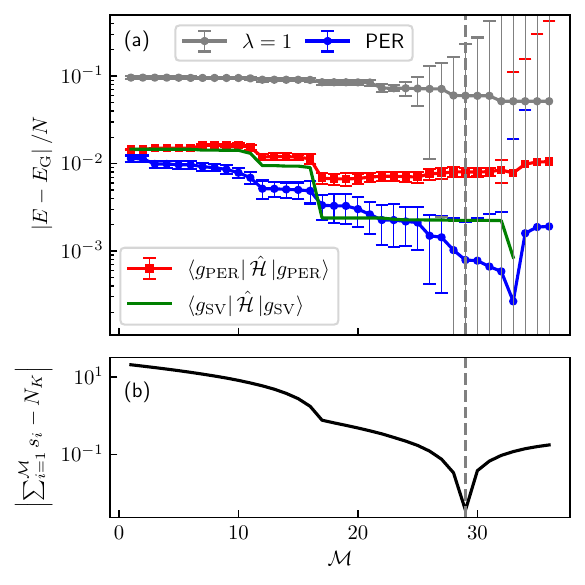}
    \caption{\textbf{PIQAE results on \texttt{fake\_guadalupe} using PER quantum error mitigation.} (a) Energy error per site $\varepsilon=|\Tilde{E}_g-E_G|/N$ (grey dots) as a function of subspace dimension $\mathcal{M}$ for a 16-site MFIM on the \texttt{fake\_guadalupe} backend. We use an $L=1$ HVA ansatz for the zeroth moment state and an expansion moment $K=1$. PIQAE calculations using $\Tilde{H}_{ij}$ and $\Tilde{S}_{ij}$ obtained from PER (blue dots) yield substantially reduced error. For comparison, we include the error $ |\avihi{g_{\text{PER}}}{\h}-E_G|/N$ (red squares) of the exact energy of the FGKS ground state $\ket{g}_{\text{PER}}$ using PER, and the error $|\avihi{g_{\text{SV}}}{\h}-E_G|/N$ (green line) derived from an overlap matrix obtained from statevector simulations. 
    (b) Absolute difference $|\sum_{i=1}^{\mathcal{M}} s_i - N_K|$ of the (sorted) eigenvalues and dimension of the overlap matrix $\mathcal{S}$ as a function of $\mathcal{M}$, demonstrating $\mathcal{M}_\text{o} = 29$. 
    We set $h_x= -1$ and $h_z = 0.5$ for the model. The error bars in (a) are obtained in the same way as in Fig.~\ref{fig:qasm}.
    }
    \label{fig:PER16}
\end{figure}

We performed PNT and PER calculations on the \texttt{fake\_guadalupe} backend for the $L=16$ MFIM, utilizing a lattice matching the native qubit connectivity of the backend. Similar to Sec.~\ref{qpu}, we used an $L=1$ layer HVA ansatz at optimized angles obtained from statevector simulations with the $K=1$ FGKS expansion. For PNT, we considered a total of 150 samples for the Pauli twirl consisting of 50 samples for pair-fidelity and 100 single-fidelity. We run circuits of varying depths ([2,4,8,16]) with 1000 shots each to ensure the proper diagonalization of the noise channel. This process generated a Pauli noise model necessary for PER, where we sampled from the partial inverse of the noise model and Pauli twirl. For PER, all 83 commuting Pauli groups are measured with 1000 samples each for five noise strengths $\lambda\in[0,0.25,0.5,0.75,1,1.5]$, evaluated with 1000 shots each. This step in the procedure is the most resource-intensive. We then extrapolate each Pauli expectation value to zero noise using the PER results obtained for different noise strengths $\lambda$ for all measured Pauli groups. These ZNE estimated expectations determine $\mathcal{H}_{ij}$ and $\mathcal{S}_{ij}$, which are used to perform the PIQAE calculation. Note that we solve a single generalized eigenvalue problem here to obtain $\tilde{E}_g$, $\ket{g}$, and $\varepsilon$ and the results are shown in Fig.~\figref{fig:PER16}. The average error with PER mitigated calculations decreased by an order of magnitude between $\mathcal{M}=1$ and the optimal subspace dimension $\mathcal{M}_{\text{o}}=22$. 
In comparison to results for $\lambda=1$ (grey points), indicating the original noise on \texttt{fake\_guadalupe}, we achieved a significant two-order-of-magnitude error reduction up to $\mathcal M=\mathcal{M}_{\text{o}}$. It is important to emphasize that the ZNE results with PER closely approximate the exact calculations, wherein $\ket{g}$ is derived by diagonalizing the overlap matrix obtained from statevector simulations. This is illustrated by the green line in Fig.~\figref{fig:PER16}{(a)}. When comparing the results depicted in Fig.~\figref{fig:qpu16}{(a)} with those in Fig.~\figref{fig:PER16}{(a)}, it becomes evident that PER proves more effective in mitigating device errors, and performing PER on quantum hardware is a promising next step. 

\section{Conclusion}
In this paper we highlight the generic idea of improving the accuracy of ground state energy calculations by synergistically integrating VQE with quantum subspace expansion (VQESE). As a specific implementation, we presented a detailed study of the paired iterative quantum-assisted eigensolver (PIQAE) in the presence of hardware noise, providing a benchmark towards near-term quantum computing applications. We have shown that by balancing the depth of the VQE ansatz $L$, which sets the zeroth moment state $\ket{\Psi}$, with the dimension of the subspace, set by $K$, one can tailor the PIQAE method according to the available QPU and CPU resources. We have demonstrated a significant enhancement of the accuracy of the ground state energy on \texttt{ibm\_guadalupe} by about one order of magnitude compared with HVA.

Using statevector simulations of the 1D and 2D TFIM and MFIM, we have established that the convergence of the energy with the moment $K$ occurs more rapidly for larger $L$. Thus, one can choose the PQC depth $L$ and the subspace expansion moment $K$ in order to achieve accurate ground state estimations, which are compatible with available resources. We also analyzed the impact of shot noise and proposed a criterion based on the trace of the overlap matrix to determine the optimal truncated subspace dimension $\mathcal{M}_\text{o}$. We used this criterion to perform PIQAE calculations on IBM hardware and have simulated a $5$-site MFIM on \texttt{ibmq\_quito} and a $16$-site MFIM on \texttt{ibmq\_guadalupe}. For the system with fewer qubits, we have obtained reasonably accurate results with an energy error per site of $\varepsilon<0.01$, corresponding to relative error $\varepsilon_{\text{rel}} < 0.7\%$.
For the 16-site system, we employed standard error suppression and mitigation techniques including ZNE to achieve an average error per site of $\varepsilon \approx 0.05$ ($\varepsilon_{\text{rel}} \approx 3\%$). Finally, we have demonstrated that using a more controlled quantum error mitigation technique such as PER can significantly improve the energy estimates and we report an improvement by two orders of magnitude to $\varepsilon_{\text{PER}} = 0.001$ ($\varepsilon_{\text{rel}} \approx 0.06\%$) for a 16-site MFIM on \texttt{fake\_guadalupe}. 

Our study is a detailed benchmark of the performance and robustness of QSE approaches in the presence of hardware noise and thus lays the groundwork for future applications of this method on QPUs to larger models that are no longer accessible via ED. When combined with controlled quantum error mitigation methods, our work suggests that PIQAE is a viable candidate to perform ground state calculations on such large systems to achieve quantum utility before fault tolerance, adding to the prospect for utility in quantum dynamics simulations~\cite{kimEvidenceUtilityQuantum2023}. Concretely, for PIQAE calculations of a $N = 127$ heavy-hex spin lattice model on \texttt{ibm\_sherbrooke} with expansion moment $K=1$, we estimate the number of distinct measurement circuits to be smaller than $8\times 127+3=1019$ according to the upper bound given by Eq.~\eqref{eq: ub}. Combined with controlled quantum error mitigation methods, such as those used here~\cite{mari2021extendingqp, McDonough-AutomatedPER-2022} and demonstrated in Ref.~\cite{kimEvidenceUtilityQuantum2023}, we expect that PIQAE can provide accurate estimates of the ground state energy and other ground state observables for large model sizes that are not accessible by classical computational approaches.


\section*{Acknowledgements}
Y. Y. acknowledges useful discussions with H. Barona. 
P.S. and P.P.O. thank Benjamin McDonough for discussions related to this work.
This work was supported by the U.S. Department of Energy (DOE), Office of Science, Basic Energy Sciences, Materials Science and Engineering Division, including the grant of computer time at the National Energy Research Scientific Computing Center (NERSC) in Berkeley, California. The research was performed at the Ames National Laboratory, which is operated for the U.S. DOE by Iowa State University under Contract No. DE-AC02-07CH11358. 
The calculation and analysis of 16-qubit model on QPU were supported by the U.S. Department of Energy, Office of Science, National Quantum Information Science Research Centers, Co-design Center for Quantum Advantage (C2QA) under contract number DE-SC0012704.

\bibliography{ref}

\end{document}